\documentclass{article}

\usepackage{epsfig}
\title{Weak Interactions Effect on the P-N Mass Splitting and the
Principle of Equivalence} 
\author{N. Chamoun\\ \emph{Department 
of Physics, HIAST, P.O. Box 31983, Damascus, Syria.}\thanks{Part of work has been done at Departamento de F\'{\i}sica, Universidad
Nacional de La Plata, c.c. 67, 1900 La Plata, Argentina.} \\ 
H. Vucetich\\
\emph{Instituto de  F\'{\i}sica, UNAM, M\'exico},
\thanks{On leave of absence from Observatorio Astron\'{o}mico,
Universidad Nacional de La Plata, Paseo del Bosque S$/$N, CP 1900 La
Plata, Argentina.}}
\date{}

\begin{document}
\maketitle
\begin{abstract}
The weak interaction contribution to the proton neutron mass
difference is computed using a generalization of Cottingham's formula.
When included in the analysis of the E\"otv\"os experiment, this
contribution reduces the bound on a possible weak interactions
violation to the equivalence principle by one order of magnitude.
\end{abstract}

\section{Introduction}
\label{Introduction}

    The Principle of Equivalence is the physical basis of General
Relativity. It loosely states that any freely falling reference frame
is locally equivalent to an inertial reference frame
\cite{Einstein1911}. This is a very strong statement: its unrestricted
validity leads to General Relativity as the unique theory for the
gravitational field \cite{Weinberg72} and experimental tests of its
consequences probe deeply the structure of gravitation.

	The validity of the equivalence principle has been studied in
the weak interaction sector, both from neutrino oscillations
\cite{Gasperini,Mann96} and $K_0-\bar{K}_0$ physics
\cite{Kenyon,Hambye}. In either of the leptonic and mesonic sectors of the
standard model, the bounds found for the breakdown of the equivalence
principle are much smaller than those found in the baryonic
sector. However, it is important to study the baryonic sector since the equivalence
principle may be well satisfied for ultrarelativistic neutrinos and kaons
while being violated by the weakly interacting nonrelativistic baryons.

	In this case, we must turn to one of the consequences of the equivalence principle, namely
the Universality of Free Fall (UFF), which states that the world line
of a test body submerged in a gravitational field is independent of its
composition and structure \cite{Will81}. In order to clarify the
former statement, let us write the Lagrangian of a test body in the
non-relativistic approximation in the form \cite{Haughan79}:
\begin{equation}
L = -m_R c^2 + \frac{1}{2} m_I {\bf v}^2 - m_P \phi({\bf x}) +
    O\left(\frac{v^4}{c^4}\right)
                        \label{Lagr-NR}
\end{equation}
where ${\bf v}$ and ${\bf x}$ are the velocity and the coordinate of
the center of mass of the test body, $\phi$ is the gravitational
potential and the parameters $m_R$ , $m_I$ and $m_P$ are called,
respectively, the rest, inertial and passive gravitational masses for
the test body.  UFF implies the equality of inertial and passive
gravitational masses:
\begin{equation}
m_I = m_P                   \label{mI=mP}
\end{equation}
while Local Lorentz Invariance (LLI) ---another consequence of the
equivalence principle--- implies the additional equality:
\begin{equation}
m_R = m_I                   \label{mR=mI}
\end{equation}

    UFF, among the consequences of the equivalence principle, is
one of the strongest tests of its validity. For instance, it has
been shown that sufficiently sensitive related experiments can
provide strict tests on superstring theories (see, eg.
\cite{DamPoly94}) or Kaluza-Klein theories (eg. \cite{OverWess}),
thus exhibiting the presence of ``new physics''. Indeed, the STEP
satellite experiment \cite{STEP93,STEP94} will improve these tests
sensitivity by about six orders of magnitude.

One of the profound consequences of the equivalence principle is that
all forms of non--gravitational energy, since they contribute to the
inertial mass, should couple in the same way to the gravitational
field. Any violation of UFF should break equation (\ref{mI=mP}) and
the difference between inertial and passive gravitational mass of a
test body could be expressed via phenomenological parameters $\Gamma_t$
specific to each type of interactions $t$ reflecting its degree of
violation to the equivalence principle:
\begin{equation}
m_P - m_I = \delta m = - \sum_t \Gamma_t E^t  \label{delta-m}
\end{equation}
where the nuclear binding energies $E^t$ can be estimated using the
semiempirical mass formula \cite{Eis78} or, in the case of weak
interactions, a suitable generalization \cite{Hau76,Fish85}.  In
principle, the parameters $\Gamma_t$ are measured in E\"otv\"os-like
experiments where they are fitted to data, but they can also be
predicted in some given theories of gravitation, thus providing a
sensitive test of such theories.

	E\"otv\"os experiments
\cite{Will81,Hughes92,Vucetich96,DamourEPC97} set an upper limit on
the difference of acceleration in a gravitational field for different
materials and so impose upper bounds on the violation parameters
$\Gamma_t$. While most published estimates, taking into account only
the binding energy contribution to the nucleus mass, show that strong
and electromagnetic interactions obey the equivalence principle to an
accuracy better than $10^{-8}$ \cite{Will81,Vucetich96}, the upper
bound on any violation of the equivalence principle by the weak
interactions is much higher ($10^{-2}$) \cite{Will81,Vucetich96}. This
is not only due to the tiny contribution of weak interactions to the
total mass but also largely because the binding energy per nucleon due
to weak interactions is a very slowly varying function across the
periodic table which then leads to a large cancellation in the
analysis of E\"otv\"os experiments \cite{Hau76,Fish85}.  Although the
weak interactions sensitivity can be improved by comparing elements
which are as far apart as is possible in the periodic table, this slow
variation will destroy the accuracy obtained in any experimental test
of UFF.

	 In order to examine further the present accuracy of
E\"otv\"os experiments with respect to weak interactions, one should
include the individual nucleons contribution to the nucleus mass since
it changes much faster along the periodic table.  There has not been,
to our knowledge, a study of the weak interactions effect within
nucleons in the analysis of E\"otv\"os experiments and the object of
this article is to provide just such a study. We shall evaluate
the proton-neutron mass difference due to weak interactions and
reassess the E\"otv\"os experiments results.

	There is a model-independent approach to the weak contribution
to the nucleon mass difference consisting of the development of a sum
rule that gives the nucleon self mass in terms of observable
quantities. We shall call this approach the generalized Cottingham's
formula since it was first done by Cottingham for the electromagnetic
interactions \cite{Cotting63} . This sum rule is a rigorous
model-independent way for computing the proton-neutron mass
difference. We describe very briefly this approach in section
\ref{p-n-primer} while we develop the generalized Cottingham's formula
corresponding to the weak interactions in section \ref{analysis}.
In Section \ref{Discussion} we implement the weak p-n mass splitting
result in a re-analysis of E\"otv\"os experiments' results and find
that they lead to an improved upper bound in that weak interactions
violation of the equivalence principle is less than $10^{-3}$.

\section{The proton-neutron mass difference}
\label{p-n-primer}

	One of the most interesting results in basic quantum field
theory is that the proton-neutron mass difference is finite
and can be computed, in principle, from experimental data. The method
is due to Cottingham \cite{Cotting63} and has been generalized to
strong interactions \cite{Epele91,Christ91}. In this section, we shall
recall the main steps in the derivation of Cottingham's
formula. Detailed proofs can be found in references
\cite{Cotting63,Epele91,RMG99}. 

	To first order in the fine structure constant, the
electromagnetic contribution to the self-energy of the nucleon may be
written as:
\begin{equation}
\Delta M^{em}_N = \frac{ie^2}{2(2\pi)^4} 
	\int d^4q G_{em}^{\mu\nu}(q^2) T^{em, N}_{\mu\nu} ({\bf q},
	q_0)	\label{em-self-0}
\end{equation}
where $G^{\mu\nu}_{em} = \eta^{\mu\nu}/q^2$ is the photon propagator
and $T^{em, N}_{\mu\nu} ({\bf q}, q_0)$ is the Compton scattering
amplitude of a virtual photon with momentum $q$ by a nucleon $N$ at
rest. In the Born approximation this amplitude reduces to:
\begin{eqnarray}
T^{em, N}_{\mu\nu} ({\bf q}, q_0) &=& 
	\frac{(2\pi)^4}{2} \frac{4Mq^2}{q^4 - 4M^2{q^0}^2} 
	\left(1 + \frac{q^2}{2M^2}\right)\nonumber\\
& &	\sum_{\rm spin} [\langle N\mid J^{em}_\mu (0) \mid N'\rangle 
	\langle N'\mid J^{em}_\nu (0) \mid N\rangle + 
	\mu \leftrightarrow \nu]		\label{Compton-0}
\end{eqnarray}
where $M$ is the mass of the nucleon $N$ at rest, $N'$ indicates a nucleon 
with four-momentum $({\bf q}, q_0 + M)$ and the sum is over both its 
spin states.

	In the same approximation, the electromagnetic current matrix 
elements between two nucleons of momentum $p$, $p+q$ and spin $\alpha$
and $\alpha '$ can be expressed in the form:
\begin{eqnarray}
\lefteqn{\langle N(p,\alpha)\mid J^{em}_\mu (0) \mid N'(p+q,\alpha
		')\rangle  =}
		\nonumber\\
	& &\bar{u}^{(\alpha)}(p) [ F^N_1(q^2) \gamma_\mu + 
		i F^N_2(q^2) \sigma_{\mu\nu} q^\nu
			 ] u^{(\alpha ')}(p + q)	\label{Def-DPff}
\end{eqnarray}
where $u(p)$ are Dirac spinors and $F_1, F_2$ are the Dirac and Pauli
form factors of the nucleon.

Plugging (\ref{Def-DPff}) into (\ref{Compton-0}) and doing a Wick
rotation, one can get, after some algebra, the expression for the
electromagnetic nucleon self energy:
\begin{eqnarray}
\Delta M^{em}_N &=& -\frac{1}{\pi} \int^\infty_0 \frac{q\,dq}{q^2}
	\int^q_0 d\nu \sqrt{q^2 - \nu^2} 
		\frac{4M q^2}{q^4 + 4M^2\nu^2}  \nonumber\\
	& & \left[3q^2 f_1(q^2) - (q^2 + 2\nu^2) f_2(q^2) \right]
		\label{em-self-1}
\end{eqnarray}
where the quantities $f_1(q^2), f_2(q^2)$ can be written in terms of the
electromagnetic Sachs form factors $G^N_{E,M}$ of the nucleon:
\begin{eqnarray}
f_1(q^2) &=& \frac{\alpha}{\pi}
	\frac{G^2_M(q^2) - G^2_E(q^2)}{q^2 + 4M^2}
					\label{Def-f1}\\
f_2(q^2) &=& \frac{\alpha}{\pi}
	\frac{q^2 G^2_M(q^2) + 4M^2 G^2_E(q^2)}%
		{q^2(q^2 + 4M^2)} 	\label{Def-f2}
\end{eqnarray}
while, in turn, the Sachs form factors are expressed in terms of the Dirac 
and Pauli form factors via
\begin{eqnarray}
G_E(q^2) &=& F_1(q^2) + \frac{ q^2}{4M^2} F_2(q^2)
					\label{Def-GE}\\
G_M(q^2) &=& F_1(q^2) + F_2(q^2)	\label{Def-GM}
\end{eqnarray}

	The Sachs form factors, which can be measured from $e$-nucleon
scattering data, have a simple physical interpretation in that
they are closely related to the Fourier transforms of the nucleon
charge and magnetic moment densities respectively.

	Equation (\ref{em-self-1}) with (\ref{Def-f1}) and
(\ref{Def-f2}) is the celebrated Cottingham's formula. It expresses
the electromagnetic contribution to the nucleon self mass as a
weighted integral on the observable form factors and the results are
finite, due to the fast decrease of the measured $G_i$. The
electromagnetic contribution to the proton-neutron mass difference is
obtained by subtracting the two electromagnetic self masses of the
proton and the neutron $\Delta M^{em}_{p-n} = \Delta M^{em}_{p} -
\Delta M^{em}_{n}$.  Using the``Galster parameterization''
\cite{Mus94,Gal71} for the electromagnetic form factors, a numerical
integration results in:
\begin{equation}
\left( \frac{ M^n - M^p }{M}\right)^{em} = - 8.3\times10^{-4}
					\label{Delta-em-pn}
\end{equation}
which amounts to a nucleon mass difference of $-0.79$ MeV making the proton 
heavier than the neutron.

	In the same way, the ``strong'' contribution to proton-neutron
mass difference can be traced to $\rho-\omega$ mixing \cite{Epele91}
or computed assuming certain models such as Skyrme models
\cite{Jain89,Epele89}, chiral solitonic models \cite{Park-Rho} and
Sigma models \cite{Clement92}. In \cite{Epele91}, an equation of the
form (\ref{em-self-1}) was established for the mass difference, in
terms of the strong $\rho NN$, $\omega NN$ and the $\rho-\omega$
mixing parameter $\epsilon$, with the result:
\begin{equation}
\left( \frac{ M^n - M^p }{M}\right)^{st} = 2.22\times10^{-3}
					\label{Delta-st-pn}
\end{equation}
which is equivalent to a mass difference of $2.08$ MeV. The final
result is the sum of (\ref{Delta-em-pn}) and (\ref{Delta-st-pn}):
\begin{equation}
\left( \frac{ M^n - M^p }{M}\right)^{tot} = 1.39\times10^{-3}
					\label{Delta-tot-pn}
\end{equation}
equivalent to a mass split of $1.31$ MeV in excellent agreement with
the experimental value $1.35$ MeV. A careful error analysis of these
results can be found in reference \cite{RMG99}.

	The above results are valid in the Born approximation,
i.e. the lowest order in $\alpha$ while higher order corrections to
the Cottingham formula are divergent and must be properly
renormalized. However, following \cite{GassLew} for a careful
discussion of this renormalization, we can see that the corrections to
the mass difference, which depend on the renormalization point $\mu$,
are very small and have no practical importance. This is because the
mass differences between particles belonging to the same isospin
multiplet are finite in the chiral limit $m_q = 0$ and all the
corrections introduced through counter terms are of the order of
$O(m_q/M)$, smaller than experimental errors. The same situation
occurs with respect to the breakdown of isospin symmetry and other
similar higher order effects.

\section{Analysis of the weak p-n mass splitting}
\label{analysis}

	In this section, we shall derive a weak Cottingham's formula
to express the weak p-n mass splitting value in terms of experimental
weak form factors.

Our starting point is the formula for the four-fermion interaction as
a low energy approximation to the IVB theory corresponding to exchange
of ($W^+$, $W^-$, $Z^0$) bosons:
\begin{eqnarray}
\label{start}
{\cal L}^{\rm eff}\;=\;{\cal L}^{\rm eff}_{\rm cc}+ {\cal L} ^{\rm
	eff}_{\rm nc}
	&=&\frac{-g^2}{2M_{W}^2}
J^{+}_{\mu}J^{-\mu} + \frac{-g^2}{2M_{W}^2}J^{N}_{\mu}J^{N\mu}
\end{eqnarray}
where, restricting our attention to one family of fermions, the charged
current is given by
\begin{eqnarray}
J^{+}_{\mu}\;=\;J^{+V}_{\mu}-J^{+A}_{\mu}&=&\frac{1}{2}
\sum_{f=\nu,e,u,d}\bar{f}\gamma_{\mu}(1-{\gamma}_{5})T^{-}f \nonumber\\
J^{-}_{\mu} &=& \left( J^{+}_{\mu} \right)^{\dagger}
\end{eqnarray}
and the neutral current is given by
\begin{eqnarray}
\label{neutral-current}
J^{N}_{\mu} 
\;&=&\;
\frac{1}{2}\sum_{f=\nu,e,u,d}\left[\bar{f}\gamma_{\mu}
\left( T_3-2Q\sin^2\theta_W
\right)f-\bar{f}\gamma_{\mu}\gamma_5T_3f \right] \nonumber\\
&=&J^{NV}_\mu - J^{NA}_\mu
\end{eqnarray}
where $Q$ is the charge matrix, $T_i=\frac{\sigma_i}{2}$ are the
generators of $SU(2)$ algebra, $T^{\pm}=T_1\,\pm\,iT_2$ and the vector
and axial components correspond to the $\gamma_\mu$ and
$\gamma_\mu\gamma_5$ terms respectively.  We deduce that the weak
interactions would contribute a term in the Hamiltonian of the nucleon
given by
\begin{eqnarray}
H&=&\frac{4G_F}{\sqrt{2}} J^{+}_{\mu}J^{-\mu} +
\frac{4G_F}{\sqrt{2}} J^{N}_{\mu}J^{N\mu}
			\label{Eff-W-Ham}
\end{eqnarray}
and our objective is to calculate the difference between proton and neutron
matrix elements of this operator since it gives the p-n mass splitting due
to weak interactions. 

	It should be noted that approximating ${\cal L}^{\rm eff}$ in
the form (\ref{start}) for purely hadronic interactions presumably has
large QCD corrections, which can be estimated as
$\log(m_W^2/m_\rho^2) \sim 9$ assuming $m_\rho=770\;{\rm MeV}$
to be a typical strong interaction scale \cite{Peskin}. To take into
account these effects, we shall introduce an enhancement factor ${\cal
G}$ in the Hamiltonian (\ref{Eff-W-Ham}). In \cite{Fish85} this factor
has been estimated to be ${\cal G} \sim 7$ from current algebra
considerations, and we shall use:
\begin{equation}
{\cal G} \sim 8 			\label{G-enh}
\end{equation}
as a reasonable estimate of ${\cal G}$.

Now, following the steps sketched in section \ref{p-n-primer}, we can
develop a sum rule corresponding to the weak interactions and which is
similar to Cottingham's formula.  Because of weak isospin symmetry we
can see that neither charged currents nor the axial part of the
neutral current will contribute to the neutron-proton mass
difference. Only the vector neutral current will give a nonzero
contribution for the difference. This current, however, has the same
structure as the electromagnetic current and so the assumptions
involved in the derivation of Cottingham's formula are still
valid. Indeed, following the steps in the derivation of
(\ref{em-self-1}) and noting that the term $e J^{em}_{\mu} A^{\mu}$
for the electromagnetic part of the Hamiltonian is substituted by the
term $\frac{g}{\cos\theta_W} J^{NV}_{\mu} Z^{\mu}$ for the weak
neutral vector part, one gets the similar result:
\begin{eqnarray}
\Delta M^{W-NV}_N &=& -\frac{1}{\pi} \int^\infty_0{\frac{q}{M^2}\,dq}
	\int^q_0 d\nu \sqrt{q^2 - \nu^2} 
		\frac{4M q^2}{q^4 + 4M^2\nu^2}  \nonumber\\
	& & \left[3q^2 f_1^Z(q^2) - (q^2 + 2\nu^2) f_2^Z(q^2) \right]
		\label{wk-self-1}
\end{eqnarray}
where the quantities  $f_1^Z(q^2), f_2^Z(q^2)$ are related to the
neutral weak form factors:
\begin{eqnarray}
f_1^Z(q^2) &=& \frac{\alpha_W}{\pi}
	\frac{[G^Z_M(q^2)]^2 - [G^Z_E(q^2)]^2}{q^2 + 4M^2}
					\label{Def-f1Z}\\
f_2^Z(q^2) &=& \frac{\alpha_W}{\pi}
	\frac{q^2 [G^Z_M(q^2)]^2 + 4M^2 [G^Z_E(q^2)]^2}%
		{q^2(q^2 + 4M^2)} 	\label{Def-f2Z}
\end{eqnarray}
where $M$ is the nucleon mass $\approx1$ GeV, and
\begin{equation}
\alpha_W =  \frac{\sqrt{2}G_F M^2}{\pi} = 0.463\times10^{-5} 
\end{equation}

	The sum rule (\ref{wk-self-1}) is the contribution to the self
mass of the nucleon coming from the isospin-breaking part of the weak
interaction which is, as we said above, related to the vector part of
the weak neutral current. The weak contribution to the proton-neutron
mass difference is obtained, then, by straightforward subtraction of
the proton and neutron weak neutral vector self masses $\Delta
M^{W}_{p-n} = \Delta M^{W-NV}_{p} - \Delta M^{W-NV}_{n}$.

	The weak form factors, except for isolated points, have not
been measured \cite{Meiss99}. However, using CVC, they can be related
to the electromagnetic form factors \cite{Mukh98}:
\begin{eqnarray}
G^{pZ} &=& \frac{1}{2} (G^p - G^n) - 2\sin^2\theta_W G^p -  \frac{1}{2}
	G^{sZ} 			\label{Def-GZp}\\
G^{nZ} &=& - \frac{1}{2} (G^p - G^n)  -  \frac{1}{2}
	G^{sZ} 			\label{Def-GZn}
\end{eqnarray}
where we have normalized them to the weak isospin values
$G_E^{p,nZ}(0) = t_{3L}$ and where $G^{s}$ is the contribution of the
$s$-quark sea to the weak form factor. Measurements show that this
latter quantity is very small and we shall neglect it
\cite{Meiss99}.

	The ``weak Cottingham formula'' (\ref{wk-self-1}) provides, in
principle, a model independent calculation of the proton-neutron mass
difference. The measured form factors neatly package many things that
cannot yet be computed {\em ab initio}, such as the QCD structure of
the nucleon. As discussed in \cite{GassLew}, the corrections to the
``weak Cottingham formula'' introduced by the renormalization process
should be much smaller than the rather large experimental uncertainties.

 We use the ``Galster parameterization'' \cite{Mus94,Gal71} for the
electromagnetic form factors and get the final result
\begin{equation}
\left( \frac{ M^n - M^p }{M}\right)^{W} = (-5.0 \pm 1.0)\times 10^{-9}
					\label{Delta-weak-pn}
\end{equation}
equivalent to a mass split of $-4.7 \pm 0.9$ eV. The error was
estimated from the known discrepancies of the Galster parameterization
with experiment, plus a generous allowance for the largely unknown
strange contribution.

\section{Discussion and the E\"otv\"os experiments}
\label{Discussion}

In order to see how this result can be implemented in a reanalysis of
E\"otv\"os experiments, we remind that these experiments, by measuring
the difference of acceleration $a$ for different bodies (say $A,B$)
falling freely in a gravitational field $g$, set an upper limit on the
difference in $\frac{\delta m}{m_I}$ for these bodies, where $m_I$ is
the inertial mass and $\delta m = m_P - m_I$ is the passive--inertial
mass difference, which serves to define the E\"otv\"os parameters
$\eta(A,B)$ via
\begin{eqnarray}
a_A - a_B \;\equiv\; \eta (A,B)\;g &=&
\left[ \left(\frac{\delta m}{m_I}\right)_A -
\left(\frac{\delta m}{m_I}\right)_B \right]\;g
\end{eqnarray}
Considering the mass of a nucleus with $Z$ protons, $N$ neutrons and
binding energy $B$
\begin{eqnarray}
m\left(Z,N\right) &=& Z M^p +N M^n -B
\end{eqnarray}
one then introduces the violation parameters $\Gamma_{t=S,W,E}$
corresponding to different types of interactions (strong, weak and
electromagnetic) through equation (\ref{delta-m}).
As we said earlier, the binding energy per nucleon $\bar b = B/(N+Z)$
is changing slowly across the periodic table  and one should take into
account the individual nucleons contribution to the nucleus mass
in order to refine the analysis, so we get
\begin{eqnarray}
\delta m &=& \left(\frac{N-Z}{2}\right)(\delta M^n-\delta M^p) + (N+Z)
\overline{\delta M} - \sum_{t=S,W,E}\Gamma_t E^t
\end{eqnarray}
where $\overline{\delta M} = \frac{\delta M^p+\delta M^n}{2}$ is the
individual nucleon average passive--inertial mass difference.  For
simplicity, we will assume, plausibly, that the violation parameters
are similar for the binding energy ($3^{rd}$ term) and the nucleon
mass--difference ($1^{st}$ term) above then we have
\begin{eqnarray}
\label{deltaM}
\delta m &=&
\sum_{t=S,W,E}\Gamma_t \left[
\left(\frac{N-Z}{2}\right)\left(M^n-M^p\right)^t - E^t \right]
+\left(N+Z\right)\overline{\delta M}
\end{eqnarray}
where $\left(M^n-M^p\right)^t$ is the neutron--proton mass splitting
due to interactions of type $t$.  Since $\overline{\delta
M}$,$M^n$,$M^p$ are invariant across the periodic table one can see,
considering the slow change of $\bar b$ and the fact that
$\left(\frac{N-Z}{2}\right)\left(M^n-M^p\right)$ is negligible
compared to $\left(N+Z\right)\left( \frac{M^n+M^p}{2} \right)$, that
the last term of Eq.(\ref{deltaM}) divided by $m$ is practically independent of
the nucleus nature and can be dropped altogether from the E\"otv\"os
parameters expression, so we get
\begin{eqnarray}
\eta \left(A,B\right) &=&
\left( \frac {\sum_{t=S,W,E}\Gamma_t \left[
\left(\frac{N-Z}{2}\right)\left(M^n-M^p\right)^t - E^t \right]} {m}
\right)_A 
\nonumber\\ 
&-&
\left( \frac {\sum_{t=S,W,E}\Gamma_t \left[
\left(\frac{N-Z}{2}\right)\left(M^n-M^p\right)^t - E^t \right]} {m}
\right)_B 
\label{eotvos-paramters}
\end{eqnarray}

	With the known expressions for $E^t$
\cite{Eis78,Hau76,Fish85},  the values of
$\left(M^n-M^p\right)^{S,E}$ \cite{Cotting63,Epele91} given by
equations \ref{Delta-em-pn} and \ref{Delta-st-pn} (section
\ref{p-n-primer}) and taking our result (equation \ref{Delta-weak-pn})
for $\left(M^n-M^p\right)^W$, we could compare to the experimental
$\eta \left(A,B\right)$ parameters in order to set bounds on
$\Gamma_t$ (See Table \ref{eottab}).

    The last two columns of Table (\ref{restab}) show a sample of
results obtained in that way with a least squares adjustment of
Eq.(\ref{eotvos-paramters}) to the data in Table \ref{eottab}, both
excluding and including the nucleon structure contribution.  We have not
included the QCD enhancement factor as it is quite uncertain and because
we are interested in upper bounds.  We find that while the inclusion
of individual nucleons effect does not change much the upper limit on
the strong and electromagnetic violation parameters $(1/10^8)$, it
lowers the bound on $\Gamma_W$ from $(4\times10^{-1})$ to
$(3\times10^{-2})$: an order of magnitude increase in sharpness. The
first two columns of Table \ref{restab} show the much sharper upper
bounds obtained considering that only one of the basic interactions
violates the Equivalence Principle. Again, the inclusion of nucleon
structure contribution affects only slightly $\Gamma_S$ and $\Gamma_E$
but lowers by one order of magnitude the bound on $\Gamma_W$ from 
$10^{-2}$ to $10^{-3}$.

	Also, if one includes the QCD enhancement factor ${\cal G}$
with its value from equation (\ref{G-enh}), one obtains, assuming that
only weak interactions break the equivalence principle, the upper
bound:
\begin{equation}
\left|\Gamma_W \right| < 2 \times 10^{-4} 
\end{equation}
This is two orders of magnitude tighter than previously reported
bounds on $\Gamma_W$ \cite{Will81}.
	
	As a final remark, let us observe that while proton-neutron
weak mass splitting originates in the neutral currents, the
``nuclear'' contribution of weak interactions is dominated by the
charged ones \cite{Hau76,Fish85}. Thus our present results put a
strong bound on both neutral and charged currents, although the
present accuracy of the data and the large correlations between the
$\Gamma_t$ variables preclude a meaningful separation of them. The
STEP experiment, with its larger accuracy and better cover of the
periodic table may help to put bounds on the separate
currents. 
However, this will depend on the exact choice of the test mass materials
which, up till now, does not seem to be public. We can only expect, after the launch of STEP, an
enhancement of about five or six orders of magnitude for the bounds of
Table (\ref{restab}).

	Even though we should interpret our results with caution, (see
reference \cite{Nieto91} for examples on mistaken analysis related to
the principle of equivalence) they confirm that present E\"otv\"os
experiments do test weak interactions effect with an accuracy, at
least one order of magnitude, better than previous studies.

\section*{Acknowledgements}
This work was supported in part by CONICET, Argentina.
N.C. acknowledges support from TWAS.
H. V. wishes to acknowledge helpful discussions and advice from
Drs. L. N. Epele, H. Fanchiotti, C. A. Garc{\'\i}a Canal,
R. M{\'e}ndez Galain and G. Gonz{\'a}lez.

\begin{table}[p]\centering

\begin{tabular}{|l|c|r|}
\hline\hline
Materials & $\eta(A,B)\times10^{11}$ & Reference\\
\hline
Al-Au & $1.0\pm3.0$ & \cite{RKD64}\\
Al-Pt & $0.0\pm0.1$ & \cite{Braginski72}\\
Cu-W  & $0.0\pm4.0$ & \cite{KeiFall82}\\
Be-Al & $-0.02\pm0.28$ & \cite{Su94}\\
Be-Cu & $-0.19\pm0.25$ & \cite{Su94}\\
Si/Al-Cu & $0.51\pm0.67$ &  \cite{Su94}\\
\hline
\end{tabular}
\caption{Results of the  E{\"o}tv{\"o}s  experiment}
\label{eottab}

\end{table}

\begin{table}[p]\centering
\begin{tabular}{|l|c|c|c|c|}
\hline
 & { $\Delta M^{n-p} = 0$} &
    { $\Delta M^{n-p}_{Ct}$} &
    { $\Delta M^{n-p} = 0$} &{ $\Delta M^{n-p}_{Ct}$}\\
\hline
$\Gamma^S$  & $1.0\times10^{-9}$ & $1.1\times10^{-9}$ 
	& $1.2\times10^{-8}$ & $9.6\times10^{-9}$\\
$\Gamma^E$  & $1.2\times10^{-9}$ & $1.2\times10^{-9}$ 
	& $2.8\times10^{-8}$ & $1.4\times10^{-8}$\\
$\Gamma^W$  & $2.8\times10^{-2}$ & $ 1.0\times10^{-3}$
	& $4.0\times10^{-1}$ & $3.3\times10^{-2}$\\
\hline
\end{tabular}
\caption[Upper bounds for the  UFF violation parameters]{Upper bounds
for the  UFF violation parameters. The first two columns show  the
upper bounds obtained assuming that a single interaction breaks the
equivalence principle. The first column ($\Delta M = 0$) excludes 
the nucleon structure contribution while the second column ($\Delta M_{Ct}$)
includes it. The last two
columns show the upper bounds obtained assuming that all three
interactions break the equivalence principle with the same conventions for 
$\Delta M$.} 
\label{restab}
\end{table}
\end{document}